# LED illumination faceted Fresnel DOEs generating perceived virtual 3D vision

Qiang Song, Yoran Eli Pigeon, Kevin Heggarty [1,*]

[1] *Telecom Bretagne, Department of Optic, IMT-Atlantique, Technopole Brest-Iroise , CS 83818,29285 BREST, France*
*\*Corresponding author: kevin.heggarty@telecom-bretagne.eu*

**An approach for the optimization and synthesis of a phase-only faceted Fresnel type diffractive optical element (FDOE) generating 3D virtual images is proposed. The FDOE is a transmissive Fresnel type DOE array, which produces the perception of a customized floating 3D virtual image behind the FDOE when illuminated with a divergent monochromatic Light Emitter Diode (LED) source. Each DOE unit of the FDOE is optimized with a modified iterative Fourier transform algorithm (M-IFTA). Each unit of the FDOE locally deflects the incident light to the same position to form a designated view of the target plane. The FDOE is fabricated using our home-built parallel write photo-lithography machine. Numerical simulations and optical experiments are performed to verify the proposed design method. Photos of the generated image are also presented. This work can find applications in optical security, anti-counterfeiting, and holographic display.**

## 1. INTRODUCTION

The creation of light fields giving a viewer the perception of a 3D object floating in space (with all or most depth cues present) has attracted extensive research attention in both academia and industry for several years. The major application is for the real-time display of varying 3D scenes (3D video) without the need for the viewer to wear 3D glasses or headgear [1-11]. Another important application is the development of attractive and novel security holograms for example for anti-fraud document protection.

There are two main kinds of 3D holographic display: one is based on the original holography technique, using light sensitive materials to record the optical interference fringes with an optical interferometer system .This requirement makes it difficult to produce nonphysical or virtual objects, and to address cost effective mass production. The other main approach uses computer generated holograms (CGH), their advantage compared to optically recorded hologram is that they do not require the interferometer system. This enables mathematical models of diffraction and interference processes to be used with computer programming. CGH based holographic 3D displays have made significant progress recently. Usually, such 3D holographic displays are obtained with CGH loaded onto a spatial light modulator (SLM).The space-band product of the SLM is still the current major limitation [12]. The viewing angle of the reconstructed 3D image is also narrow due to the SLM pixel size limit [13].

To address these problems, a spatial division multiplexing method was proposed to enlarge the viewing angle with multiple SLMs. But, this makes the system expensive and complicated [14-15]. Similarly, a time division multiplexing method was also proposed, however, a high refresh rate SLM is required [16]. Meanwhile, the large volume of the holographic system based on SLMs is another restriction. With these factors, the practical application is limited greatly, especially in consumer photonics community such as optical security, anti-counterfeiting banknote and ID card.

Surface relief Diffractive optical elements (DOE) can be an effective method to address the problems mentioned above, and are a good choice up to now due to their small dimensions and ease of replication. DOE currently attract researcher's significant interest, fueled by their extensive applications in for example beam shaping, structured illumination for face recognition, optical tweezers[17], holographic display[18]. Generally, the pixel pitch of a DOE surface only depends on the performance of lithography machine used to fabricate the device. Mathematically, the computation of the DOE to realize the 3D vision effect is an inverse problem. Several research papers on this aspect have appeared in recent years. Anton Goncharsky and his co-authors proposed a subwavelength grating array method using electron-beam lithography (EBL) technology that can generate visual 2D and 3D images on the surface of the DOE plane [19-20]. This can give observer an impressive 3D vision effect. However, the diffraction efficiency of the gratings is generally low, and the higher orders can produce crosstalk. Multilevel gratings or blazed gratings are an effective way to improve the diffraction efficiency, but the fabrication process is very complicated with EBL. Later, the same team used a similar method to realize 360 degree 3D image with grating arrays fabricated on a cylindrical surface [21]. A similar grating array based DOE has also been fabricated with a laser interference method [22]. However, the resolution and image quality are both limited. In another related work uses a similar approach to design the grating's parameters, the DOE is fabricated by a maskless photolithography machine, which can be more efficient [23]. Maskless photolithography is an effective technique for fabricating complicated micro relief structures, although the resolution is lower than EBL. However, in most applications, the resolution of photo-lithography is sufficient.

The above mentioned method of creating 3D images with DOEs mainly focuses on grating based methods. Unlike these methods, in this work we propose another design approach to generate 3D floating virtual image with a LED source, named faceted pure phase Fresnel DOEs. LED sources are used because they are less dangerous for vision perception than lasers, and are more convenient as no beam expanding optics are required. On the surface of our FDOE, each DOE facet diffracts incident light to generate different virtual viewing angle images toward to the same observer position, it therefore belongs to the off-axis type DOE. To get high image quality, a modified phase retrieval or IFTA algorithm is used to optimize each DOE facet. After design, the FDOE is fabricated with our home-built gray scale parallel photolithography machine. When the FDOE is illuminated by a divergent LED source, a 3D floating virtual image is formed behind the FDOE; an observer can see different views of the virtual 3D object hanging in the air. The configuration is described in detail in Section 2. The design procedure is presented in Section 3. Fabrication and experiment results are presented in Section 4, results are discussed in Section 5. Finally, we conclude our work in Section 6.

## 2. PRINCIPLE AND CONFIGURATION

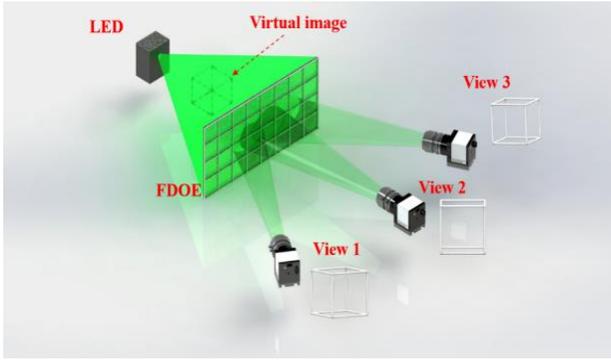

**Fig. 1.** The configuration of the LED source and observer.

Figure.1 describes the layout of observations which comprises a LED source and FDOE. The LED source is behind the FDOE. For producing the 3D desired visual effect, we propose an approach based on faceted Fresnel DOEs. In such an approach the overall diffractive structure is separated into a matrix of individual Fresnel DOEs. Each facet diffracts to generate a view of the same basic 3D object, but seen from a slightly different angle. By carefully optimizing the angles for each facet, when an observer moves their eye from side to side at the FDOE plane, they observe the different DOE facets, generating the slightly different angular views of the 3D object. The overall effect is that of observing a real 3D object. To make sure the observer has the impression that each view comes from the same object in the same position, the different DOE facets must generate their reconstructed images with the correct off-axis offset. The lateral displacement of each reconstructed image should be calculated carefully. The basic calculation principle is shown below in Fig.2.

The global coordinates of the FDOE plane are denoted by ($x_0$, $y_0$), and the local coordinates of the (m, n)-th DOE are denoted by ($x_m$, $y_n$). The distance between the FDOE plane and the virtual image plane is denoted by -z, where z is positive value, the distance is negative. The complex

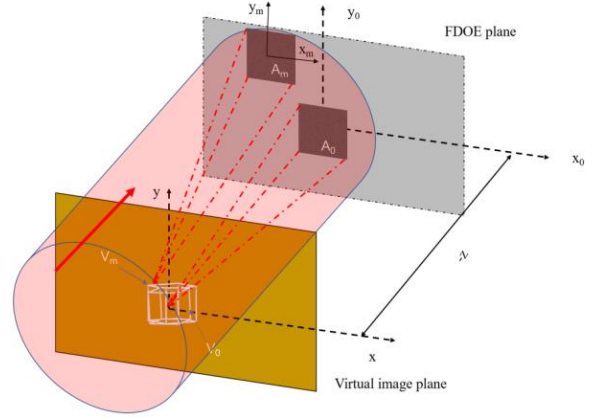

**Fig. 2.** Principle of proposed FDOE calculation process.

amplitude distribution of (m, n)-th DOE on FDOE plane can be written as below:

$$H_{m,n}(x_m, y_n) = A_{source}\exp[j\varphi_{m,n}] \quad (1)$$

Where $A_{source}$ and $\varphi_m$ are the amplitude of the incident light and, the m-th DOE's phase distribution, respectively. According to Fresnel diffraction theory, the diffracted field distribution of the (0,0)-th DOE in the virtual image plane can be calculated by

$$U_0(x, y) = \frac{\exp(-jkz)}{-j\lambda z}\exp\left[j\frac{k}{-2z}(x^2+y^2)\right]$$

$$\times \mathcal{F}_{2D}\{H_0(x_0,y_0)\exp\left(\frac{jk}{-2z}(x_0^2+y_0^2)\right)\} \quad (2)$$

In eq.(2), k=2π/λ is the wave number in free space, λ is the design wavelength, and $\mathcal{F}_{2D}$ represents the two dimensional Fourier transform. In order to generate the reconstructed virtual image of the (m, n)-th DOE toward to the same position as that of the (0,0)-th DOE, eq.(2) should be modified. See in Fig.2, $x_m$ and $y_n$ can be expressed as

$$x_m = x_0 + mP_{DOE} \quad (3)$$
$$y_n = y_0 + nP_{DOE} \quad (4)$$

In eq.(3) and eq.(4), $P_{DOE}$ represents the cell size of FDOE, m and n are the displacement along $x_0$ axis and $y_0$ axis, respectively. According to Fresnel diffraction theory, the complex amplitude distribution $U_{m,n}(x, y)$ in the virtual image plane contributed by the wavefront from the (m, n)th DOE can be calculated as

$$U_{m,n}(x,y) = \frac{\exp(-jkz)}{-j\lambda z}\iint H_{m,n}e^{\frac{jk}{-2z}[(x-x_m)^2+(y-y_n)^2]}dx_m dy_n \quad (5)$$

Substituting eq.(3) and eq.(4) into eq.(5), eq.(5) becomes

$$U_{m,n}(x,y) = \frac{e^{-jkz}}{-j\lambda z}e^{\left[\frac{jk}{-2z}(x^2+y^2)\right]} \times e^{\frac{jk(mP_{DOE}x+nP_{DOE}y)}{z}}$$
$$\times \iint H_{m,n}(x_m,y_n)e^{\frac{jk}{-2z}[(x_0+mP_{DOE})^2+(y_0+nP_{DOE})^2]}e^{\frac{jk}{-2z}(xx_0+yy_0)}dx_0 dy_0 \quad (6)$$

Where, $f_x$, $f_y$ are defined as

$$f_x = \frac{x}{\lambda z}, \quad f_y = \frac{y}{\lambda z} \quad (7)$$

With the 2D Fourier transform, eq.(6) can be simplified as

$$U_{m,n}(x,y) = \frac{e^{-jkz}}{-j\lambda z} e^{\left[\frac{jk}{-2z}(x^2+y^2)\right]} \times e^{\frac{jk(mP_{DOE}x+nP_{DOE}y)}{z}}$$
$$\times \mathcal{F}_{2D}\left\{H_{m,n} e^{\frac{jk}{-2z}[(x_0+mP_{DOE})^2+(y_0+nP_{DOE})^2]}\right\} \quad \textbf{(8)}$$

According to eq.(8), the diffracted field of the off-axis Fresnel DOE can still be expressed in a 2D Fourier transform format. So it can be computed quickly with the fast Fourier transform method. Sampling constraints for the Fresnel diffraction calculation should also be considered. According to the Nyquist sampling rule, the absolute value of the propagation distance z should satisfy the condition below:

$$z > \frac{N\Delta p^2}{\lambda} \quad \textbf{(9)}$$

Here, in eq.(9), N is the number of sampling points and $\Delta p$ is the DOE pitch. Generally, the complex amplitude can be obtained by inversing eq.(8) directly, the phase can then be extracted to generate the DOE surface relief structure. However this process loses amplitude information and image quality is not optimized. The double phase method [6] is an effective way to retain the whole complex amplitude information, but, two masks and a spatial filter are needed, which is inappropriate for security hologram applications. In this work, a modified iterative Fourier transform algorithm is implemented to design each off-axis Fresnel type DOE, it belongs to the IFTA family [24,25]. This method optimizes the image quality and the fabrication performance at the same time by introducing a soft quantization iterative algorithm. The fabricated phase levels are quantized, so the surface relief of the DOE should be divided into several levels. In theory, the higher the number of phase levels, the better the reconstructed image quality. But, the fabrication performance must also be considered. Usually, eight levels is a good practical compromise. The details of the design procedures are provided in Section 3.

## 3. DESIGN PROCEDURE AND NUMERICAL SIMULATIONS

### A. Design method

The flowchart of the proposed design approach is shown in Fig.3. The stack of different viewing angle target images is obtained using a 3D model. Then, each target image enters into the DOE optimization algorithm. This iterative algorithm can be divided into three parts, the so called three-part IFTA. The first step is the standard Gerchberg-Saxton(GS) algorithm, it is implemented to get the rough phase solution, the GS operation process can be considered as the transformation of signals between the input DOE plane and reconstruction plane. The GS iteration process can be described as four basic steps: (1)The process starts from the reconstructed plane. A random phase is generated with the range 0-2π as the initial phase, and then multiplied by an amplitude matrix with the current viewing angle target image to form the initial complex amplitude. (2)This is propagated to the FDOE plane using inverse Fresnel diffraction field tracing calculated by Fast Fourier Transform(FFT). (3)Note that in what follows, we assume that the DOE facets are illuminated by plane waves. Illuminating such DOE with spherical wave will simply shift the position of the observed image along the optical axis; see discussion part for more details. The amplitude of the wavefront is replaced by the intensity distribution of the source amplitude, while the wavefront phase is retained, to form the updated complex amplitude at the FDOE plane. (4)Diffracting the result back to the reconstructed plane, the phase obtained from the Fresnel diffraction propagation is combined with the corresponding view image amplitude. Finally repeat from step 1. Based on the above steps, the phase is optimized for 20 iterations, then the algorithm goes on to the second stage, where the iteration process is similar to the GS, but the constraint factor strategy on the reconstruction plane is different with GS algorithm. More details about these constraints can be found in our former works [18]. After that, the optimization process goes into the third stage; this stage is called the iterative soft quantization method. The soft quantization is a stepwise iterative quantization method. In this part, the iterative process is similar to that of the second stage, but another loop is introduced into it. This process of iteration can be divided into 10 cycles, each cycle includes 30 iterations. The soft quantization operation can be demonstrated as[4]:
First we define $\varphi_1$ as

$$\varphi_1 = (\varphi_{phase}^k)/(2\pi/Levels) \quad \textbf{(10)}$$

$$\varphi'^k_{phase} = \varphi_1, \ if \ -\frac{\varepsilon_P}{2} \leq \varphi_1 - round(\varphi_1) < \frac{\varepsilon_P}{2} \ (1 < p < 10) \quad \textbf{(11)}$$

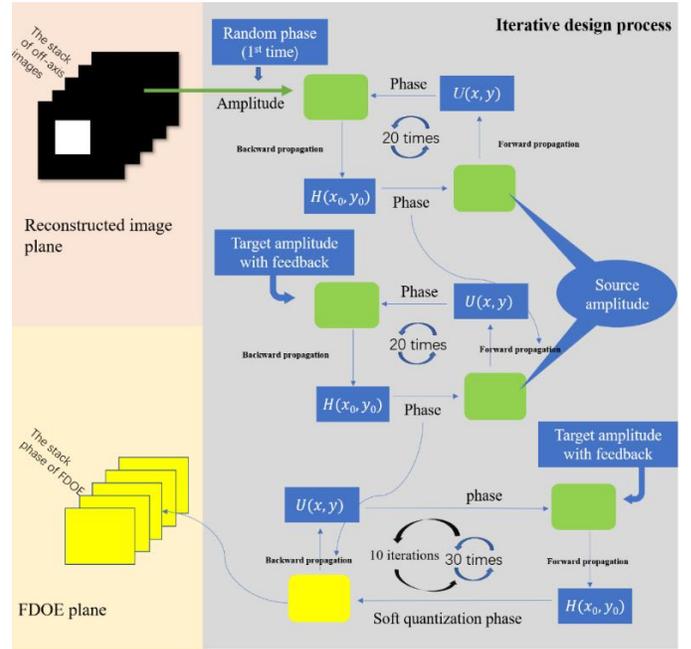

**Fig. 3.** Flowchart of the proposed iterative optimization process of FDOE

In eq.(10), $\varphi_{phase}^k$ is non-quantized phase of the k-th iteration, $\varphi'^k_{phase}$ represents the quantized phase in eq.(11). The quantity increase with the index $p$, $p$ is the current loop cycle, the $\varepsilon_P$ maximum loop is 10 in this paper. This is an empirical parameter. Here, the collection of the values in this work is

$$\varepsilon_1 = 0.15, \varepsilon_2 = 0.3, \varepsilon_3 = 0.45, \varepsilon_4 = 0.6, \varepsilon_5 = 0.7,$$
$$\varepsilon_6 = 0.8, \varepsilon_7 = 0.85, \varepsilon_8 = 0.9, \varepsilon_9 = 0.95, \varepsilon_{10} = 1 \quad \textbf{(12)}$$

The reason why we choose 30 iterations in this process will be discussed in next sub-section where the convergence behavior is described. So, in this way, the phase can be quantized into eight levels with high signal to noise ratio(SNR) reconstructed image, and adapted to the fabrication constraints. Generally speaking, the higher the number of DOE phase levels, the higher the diffraction efficiency. But, the efficiency increase for greater number of quantization level is often less than the decrease from fabrication errors. The iterative calculation stops when the error between the intensity distribution on the reconstructed plane and target image is small enough, or the maximum number of iterations is reached. The root mean square error (RMSE) can be defined as

$$RMSE = \frac{\iint (U_{target} - \gamma |U_K|)^2}{\iint U_{target}^2} \quad (13)$$

Where, $U_K$ is the reconstruction result of the k-th iteration, $\gamma$ is a scale factor, which can be written as

$$\gamma = \left(\frac{\iint U_{target}^2}{\iint |U_K|^2}\right) \quad (14)$$

When calculations of all the DOE facets are finished, the set of DOE phases are assembled to get the final FDOE.

**B. NUMERICAL SIMULATION**

For the numerical simulation, a 3D "cube" object is used to generate the different viewing angle images. In this study, we use the horizontal views. To examine the convergence behavior of the proposed optimization algorithm, one view of these image is selected as the target image, which is shown in Fig.4(a), it is the center view of the 3D object. The RMSE value against the iteration numbers is plotted, as shown in Fig.4(b).

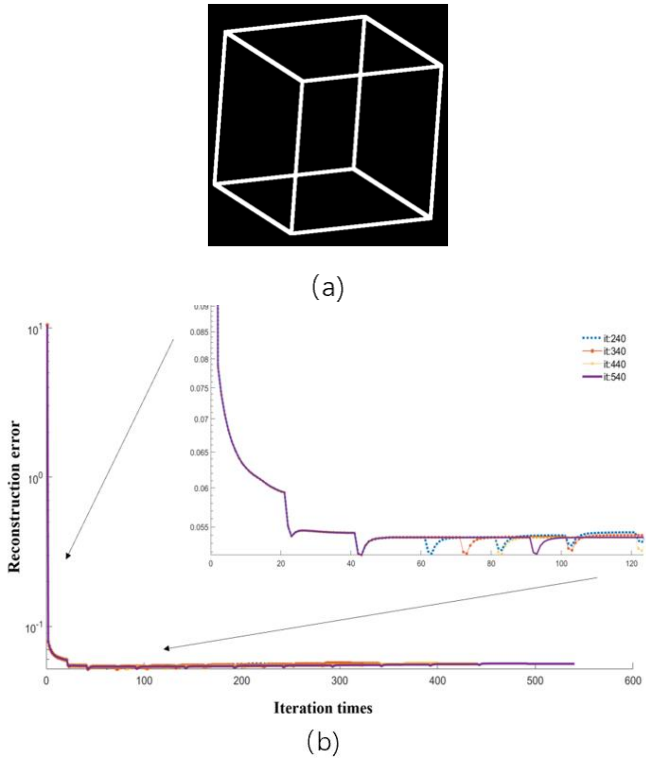

**Fig. 4.** (a) Target image for testing the algorithm's convergence behavior. (b) Curve of the reconstruction's RMSE against the iterations

In this numerical simulation work, the wavelength is 525nm, the DOE pixel pitch is 0.75um, the number of sampling points is 3600 in each DOE facet, and the distance between the FDOE plane and the virtual image is set as -120mm. Acoording to formula (13), the RMSE curves are calculated with different number of iteration in the soft quantization stage. For example, if the iterations of each loop of the soft quantization is 50, the total number of iteration is 540 (calculated as 20 (the first part) +20(the second part)+500(the third part)). See from Fig.4 (b), the RMSE is small enough when the total number of iteration is 340. The GPU is used to accelerate the calculation speed; the iteration optimization time is about 180 seconds on a computer equipped with Intel Core™ I7-8750H CPU@2.2GHz, GPU GTX1060 and the computation platform is MATLAB software. The final optimal phase is shown in Fig.5 (a) , the inset is the phase value distribution in the red circle. It is obvious the structure is eight levels. The simulated reconstructed virtual image in the virtual plane is demonstrated in Fig.5 (b), its RMSE is small enough, see in Fig.4 (b), so the reconstructed image is similar to the target image in Fig,4(a). But, the reconstructed image is a bit darker than the target image, because the calculated diffraction efficiency is about 82%, part of the energy being lost as noise. For our application, the aim is to generate horizontal 3D views. 11 off-axis Fresnel type DOE elements are used. The size of the FDOE is 29.7mm*2.7mm, it can be regarded as the system eye-box. However, this is too small in the vertical direction, an observer will not see the virtual image if their eye moves beyond this small area. The pupil size of human eye is between 6mm~8mm, so the vertical direction should be expanded to satisfy this size. One way to increase the vertical size is to increase the number of DOE sampling points, but the calculation will be heavy because the complexity is O($N^2 logN$). Here, we propose another way to increase the vertical eye-box size. The assembly is shown in Fig.6.

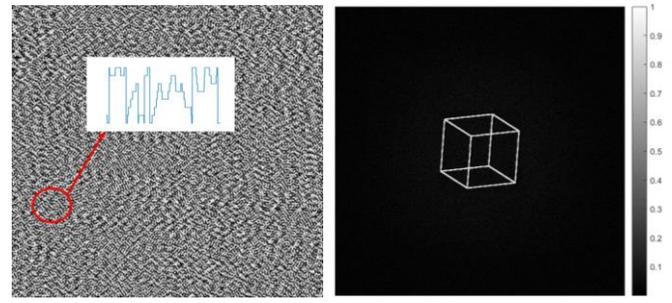

**Fig. 5.** Optimal phase map and the reconstructed result. (a) Phase map of the DOE. (b) Reconstructed virtual image.

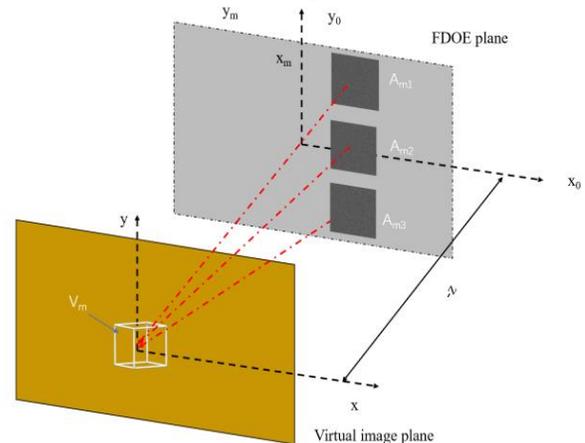

**Fig. 6.** Assembling of the FDOE.

In Fig.6, the element $A_{m0}$, $A_{m1}$ and $A_{m3}$ generate the same virtual image $V_m$. When people move their eye in vertical direction, they can still see the same virtual image at specific view. In this way, the eye-box can be expanded in vertical direction, and the final size of the FDOE is 29.7mm*8.1mm. Due to the use of divergent illumination and DOE fabrication limitations, the LED source is visible to the observer. If the FDOE are designed with an on axis target, this visibility of the LED source perturbs the desired image. To avoid this, an off-axis configuration is used.

## 4. FABRICATION AND EXPERIMENT RESULTS

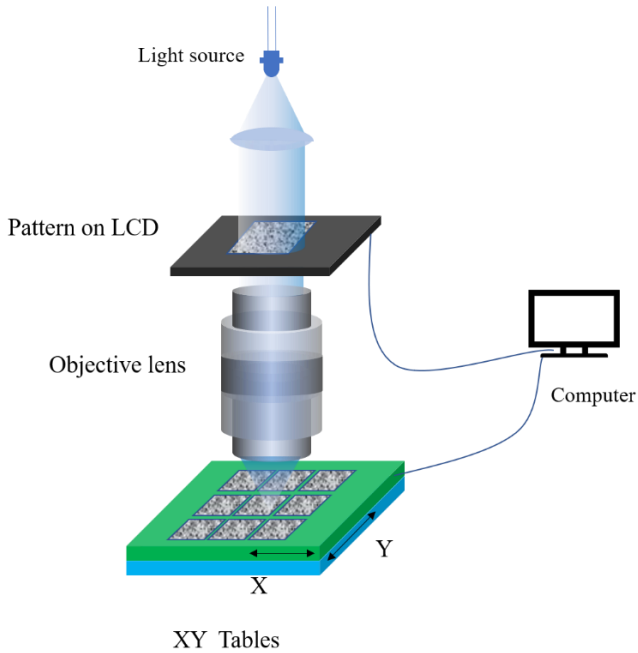

**Fig. 7.** Diagram of photo-lithography system

With the algorithm mentioned above, a FDOE was designed with 11*3 facets, The smallest pixel size of our photolithography machine is 0.75μm, but here we use 1.5μm as design pixel size. This helps minimize pixel deformation in the off-axis configuration which tends to decrease the DOE fringe size. Each DOE facet is1800*1800 pixels with pixel size 1.5μm,the total FDOE size is 29.7mm*8.1mm.The fabrication facilities we used to make FDOE are our home built direct-write photolithography machine, an updated version of what demonstrated in [26,27,28]. In Fig.7, the light source is LED based with a center wavelength of 435nm. The divergent light is first collimated and homogenized by using group of lens and diffuser. Then, the collimated and uniform light field is modulated depending on the designed phase pattern loaded on the 1920x1080 pixel Spatial Light Modulator(SLM) and de-magnified onto the surface of the photoresist on the precision moving platform through the objective lens group, The demagnification of the objective lens is 5. The projection area of SLM is small, around 4mm*3.2mm, so the FDOE is made stitching different SLM images by moving the stage in x and y direction with an accuracy of about 250nm. The fabrication process is now described briefly. There are three main steps to fabricate the FDOE. First step, the glass substrate is cleaned with ultrasonic water cleaning, and then heated for 15 minutes to dry the glass on a hot plate. The second step is the spin coating of a photoresist onto the substrate. In this work, the photoresist is the S1813 from Micro Resist Technology. The spin coating speed is chosen to give uniform photoresist layer. Etch depth is chosen to give the maximum phase difference of each cell is 1.75π. The etch depth is dependent on the wavelength λ of the incident light and the refractive index of the resist. According to the thin element approximation (TEA), the groove depth of the surface relief structure can be written as

$$d = \frac{N-1}{N}\frac{\lambda}{2(n_{material}-n_{air})} \quad (15)$$

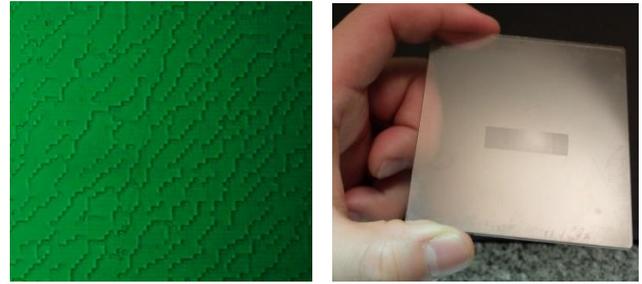

**Fig. 8.** Fragment part of FDOE and real sample. (a) Fragment of FDOE; (b) Real sample on photoresist with glass substrate.

The number of phase levels, N=8, $n_{material}$ is the refractive index of photoresist, its value is 1.66 at a wavelength of 532nm. According to this formula, the maximum depth d is about 691nm. The spin coating speed is 3200rpm, giving a photoresist thickness about 1.3μm; this depth is enough for etching the DOE. The substrate is then baked for 2 minutes at temperature of 120 degree to extract the solvent, and then cooled down to room temperature. After that, exposure is implemented using our home-built parallel direct-write photo-plotter with a lookup table (LUT) to establish a linear relation between the addressed gray phase level and the etch depth. The exposed pattern is mapped from the LCD screen to photoresist plane. The exposure time can be controlled to ensure the optimum etching depth. The third step is to develop the exposed photoresist in a chemical solution for 2 minutes. A fragment of the FDOE and final fabricated FDOE are shown in Fig.8 (a) and (b), respectively.

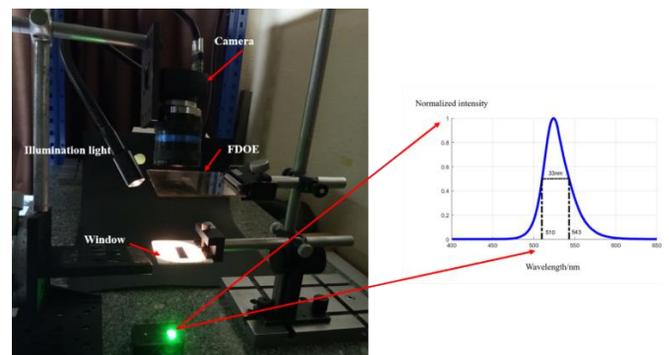

**Fig. 9.** Experiment setup for 3D floating virtual image reconstruction. Inset: spectrum of the LED source, which is obtained by Ocean Optics HR4000CG-UV-NIR, its center wavelength is 525nm.

The photo of Fig.8 (a) is captured by an optical microscope (Reichert-Jung Polyvar Met Microscope), with a 100x objective. The 3D floating virtual image can be observed with divergent LED source illuminating the FDOE. The photo of experimental set up used to capture the output views is shown in Fig.9. This system comprises a LED source, window, illumination light, FDOE and camera. A green LED source behind the FDOE is used as the illumination source for replaying the expected 3D virtual image, and the window is used for limiting the observation area, The camera is mounted on the support, which can be rotated to gain different viewing angles equivalent to those can observer sees moving his head to view the apparent 3D object. The distance between LED and FDOE is about 155mm, and position of the virtual image is about 70mm below the FDOE. The position of virtual image can be calculated according to the imaging formula, which is expressed in eq. (16).

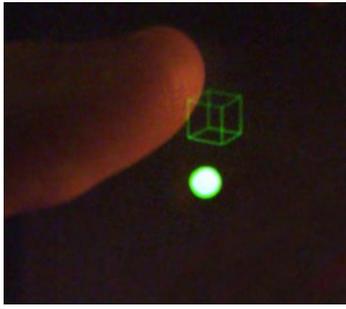

**Fig. 10.** One perspective of the 3D cube floating in the air

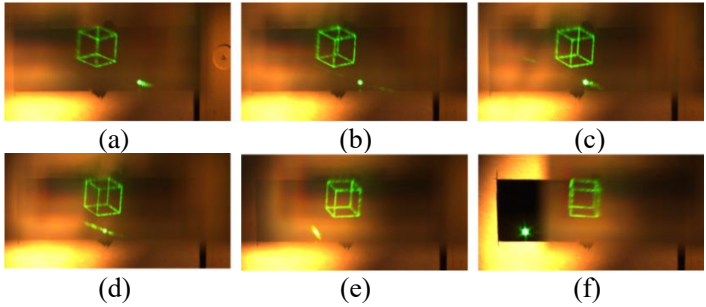

(a)          (b)          (c)

(d)          (e)          (f)

**Fig. 11.** Different virtual image views of the object "cube" with LED source. (See Visualization 1)

$$d_{image} = \frac{f_{fresnel} d_{obj}}{d_{obj} - f_{fresnel}} \quad (16)$$

In eq.(16), $f_{fresnel}$ is the focal length of the Fresnel type DOE, $d_{obj}$ is the distance between the source and the FDOE plane. With the LED light illumination, the diffracted light propagates to human eye and the 3D virtual image is perceived in the air between the DOE plane and the LED source. Fig.10 shows one perspective image of the 3D model "cube". When the hologram is optically replayed, a floating virtual image appears hanging in the air, A finger is near to the virtual image. An observer also can see a realistic 3D image by looking at the virtual object from different angle. Fig.11(a)-(f) are six reconstructed images of the FDOE, which are captured from six viewing angles by rotating the camera from left to right, see Fig.9. The different viewing images locate at the same position. The total field of view (FOV) is about 20 degree, which can be calculated with the formula $FOV = 2 \times \arctan(\frac{\lambda}{2Pitch})$. The brightness of the reconstructed images are slightly different because the exposure time of camera is slightly different for the different image.

## 5. DISCUSSION

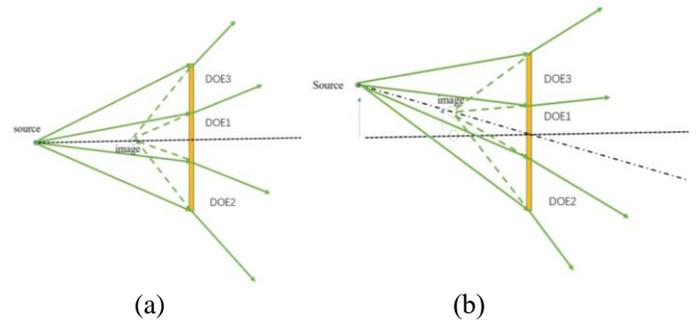

(a)          (b)

**Fig. 12.** Schematic of imaging principle, (a) LED source is on the axis, (b) LED source is off axis.

This paper has presented the design process for a diffractive structure, based on an array of computer generated off-axis Fresnel phase holograms. When illuminated by an LED readily available LED source, the overall structure gives an observer the perception of a 3D floating in space behind the hologram. The design model has been described and confirmed through both numerical simulation and experimental results. Although, this work focused on presenting the design methodology to generate perceived 3D virtual vision in the horizontal direction, the full-parallax vision can also be generated with the same approach. In Fig.6, here we just let the $A_{m1}$, $A_{m2}$ and $A_{m3}$ form the same image at the same position, if we let them form different views along the vertical direction, the observer perceives different views in the vertical direction, so the full-parallax vision can be realized in this fashion. Another important factor is the position of the LED source. In this work, the FDOE we designed is a space-invariant component, since the light propagation model used here is a linear transformation; the observed result is not sensitive to the position of LED source. Mathematically, the off-axis Fresnel type DOE can be decomposed into three linearly role as a spherical lens, here a negative lens, and the blazed grating phase is used to shift the image to the desired position. In Fig.12 (a), when the LED source is on the axis of the DOE1, the virtual image is still formed on the axis because there is no blazed grating component in the DOE1. The position of the generated virtual image can be calculated with the geometrical imaging formula, mentioned above in eq. (16). The DOE2 and DOE3 contain the blazed grating component, so, their corresponding virtual image can be deflected to the same position with DOE1. When the LED source move with respect to the original position (see Fig.12 (b)), the virtual image will move in the same direction, this process is similar to the lens imaging system, because the Fresnel lens component works as lens. In this situation, the corresponding images generated by DOE2 and DOE3 just move in the same way. But, one thing must be bear in mind, the LED source is a divergent source, and the intensity is cosine with the divergent angle. The bigger divergent angle, the weaker intensity. So, the shifting of the LED source cannot be too far. Another important factor is the spatial coherence of the LED source, which has been discussed in detail in reference [29]. According to their result, the size of the LED cannot be too big; otherwise, the reconstructed image will be not sharpness due to the low degree of spatial coherence. In this work, the LED size is 1mm*1mm, but as the distance from LED source to FDOE plane is more than 150mm, the LED source can be still regarded as 'point source'. Another important DOE parameter to be evaluated is the diffraction efficiency: the image must be sufficiently bright to be visible to the observer. Unfortunately measuring the diffraction efficiency of this type of DOE is not straightforward because the image is virtual and the illumination divergent. To obtain an estimate of the diffraction efficiency we used the setup shown in figure 13. The lens is used to convert the virtual image into a real image which is projected onto the photodetector. A diaphragm is placed between the LED source and the DOE so as to illuminate only the active DOE area. Without the DOE, the total collected light power on the detector gave an estimate of the incident light power. With the DOE present the light power in the desired image area gave an estimate of the light power in the signal. In this way we obtained a DOE diffraction efficiency estimate of 46%. The DOE zeroth order was estimated in a similar way at 20%. The remaining light losses being due mainly to Fresnel reflection (~10%) and higher diffraction orders. We are currently improving our fabrication facilities to decrease the DOE pixel size which will increase diffraction efficiency by suppressing the higher diffraction orders [30] and optimizing the setup and algorithm to make the DOE easier to fabricate. Based on values obtained with similar DOEs, we expect to be able to reduce zero order power to a few percent. However, even with the current diffraction efficiency of 46%, the observed image is already clearly visible in daylight with a readily available LED.

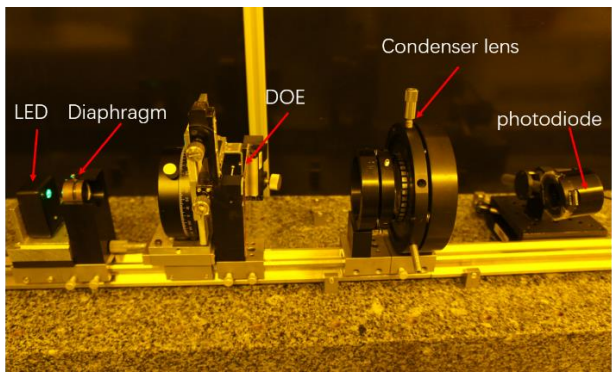

**Fig. 13.** Setup for diffraction parameters measurement.

## 6. CONCLUSION

This paper has presented a computer generated horizontal moving parallax synthetic DOE based on facet Fresnel type DOE array. The design model for the FDOE has been demonstrated through both numerical simulation and experimental results. In this work, we limit viewing of the FDOE to a horizontal displacement of the observer's eye. The 3D virtual object can be observed when the hologram is illuminated with a divergent LED source. The proposed optimization method also can be implemented to calculate a full-parallax DOE, which has been described in Section 5. The FDOE can be replicated using roll to roll nanoimprint techniques for mass production. The FDOE for creating 3D virtual object that we developed in this paper can find applications in practical holographic 3D display and anti-counterfeiting protecting of bank note and ID card.